\documentclass{aastex}
\usepackage{spr-astr-addons}
\usepackage{url}
\usepackage{amsfonts}
\usepackage{amssymb,epsfig}
\usepackage{latexsym}
\usepackage{amsmath}
\usepackage{graphicx}

\RequirePackage{color}

\begin{document}
\title{Can observational growth rate data favour the clustering dark energy models?}

\slugcomment{Not to appear in Nonlearned J., 45.}
%% Running heads
\shorttitle{Growth rate in clustering dark energy models} \shortauthors{Mehrabi et al.}

\author{Ahmad Mehrabi\altaffilmark{1}}
\altaffiltext{}{E-mail: mehrabi@basu.ac.ir}

\author{Mohammad Malekjani\altaffilmark{1}}
\altaffiltext{}{E-mail: malekjani@basu.ac.ir}

\author{Francesco Pace\altaffilmark{2}}
\altaffiltext{}{E-mail: francesco.pace@port.ac.uk}

\altaffiltext{1}{Department of Physics, Faculty of Science, Bu-Ali Sina University, Hamedan 65178, Iran}

\altaffiltext{2}{Institute of Cosmology and Gravitation, University
of Portsmouth, Dennis Sciama Building, Portsmouth, PO1 3FX, U.K.}

\begin{abstract}
Under the commonly used assumption that clumped objects can be well
described by a spherical top-hat matter density profile, we
investigate the evolution of the cosmic growth index in clustering
dark energy (CDE) scenarios on sub-horizon scales. We show that the
evolution of the growth index $\gamma(z)$ strongly depends on the
equation-of-state (EoS) parameter and on the clustering properties
of the dark energy (DE) component. Performing a $\chi^2$ analysis,
we show that CDE models have a better fit to observational growth
rate data points with respect to the concordance $\Lambda$CDM model.
We finally determine $\gamma(z)$ using an exponential
parametrization and demonstrate that the growth index in CDE models
presents large variations with cosmic redshift. In particular it is
smaller (larger) than the theoretical value for the $\Lambda$CDM
model, $\gamma_{\Lambda}\simeq0.55$, in the recent past (at the
present time).
\end{abstract}
\maketitle

\section{Introduction}
The current accelerated expansion of the universe is one of the biggest challenges in modern cosmologies.
A wide range of astronomical data, including supernova type Ia (SNeIa)
\citep{Perlmutter1997,Perlmutter1998,Perlmutter1999,Riess2004,Riess2007},
cosmic microwave background (CMB) \citep{Bennett2003,Spergel2003,Spergel2007},
large scale structures (LSS) \citep{Hawkins2003,Tegmark2004a,Cole2005}
and baryon acoustic oscillations (BAO) \citep{Eisenstein2005}
confirm the current accelerated expansion of the universe. This is a surprising
result, since in a universe dominated by matter, and therefore where gravity is attractive, we should expect the
acceleration to slow down. Within the framework of General Relativity (GR), the accelerated expansion can be
accommodated by the introduction of an unknown form of energy with sufficiently negative pressure, the so called dark
energy (DE). Observational results of the Planck satellite \citep{Planck2013_XVI} indicate that DE is $\approx 68\%$
of the total energy budget of the universe, while pressureless matter (baryons and cold dark matter) amount to
$\approx 32\%$. Hence, at the present time the dominant component is the DE with a negative equation of state (EoS)
$w_{\rm de}=p_{\rm de}/\rho_{\rm de}<-1/3$.

 A very simple and important example of fluid with negative EoS is
Einstein's cosmological constant, $\Lambda$, with $w_{\Lambda}=-1$
independent of time \citep{Weinberg1989,Sahni2000,Peebles2003}.
Although the cosmological constant is consistent with observational
data, it suffers from the fine-tuning and the cosmic coincidence
problems \citep{Weinberg1989,Sahni2000,Peebles2003}. Alternatively,
extensions and/or modifications of Einstein gravity theory, such as
$f(R)$ and $f(T)$ models, can explain the present cosmic
acceleration
\citep{Amendola2007,Hu2007,Appleby2007,Starobinsky2007,Tsujikawa2008,Cognola2008,Linder2009}.
In a general relativistic framework, DE models with equation of
state $w_{\rm de}\neq -1$ have been proposed, usually in the context
of scalar fields such as quintessence models, in order to solve or
alleviate the cosmological constant problems \citep{Peebles2003}.
Moreover, phenomenological models with time evolving EoS parameter
$w_{\rm de}(z)$ have been suggested
\citep{Linder2003,Sahni2003,Alam2003,Lazkoz2005}.

DE not only accelerates the expansion rate of the universe but also
 changes the growth rate of matter perturbations of cosmic
structures. In fact, the growth of matter perturbation $\delta_{\rm
m}$ of overdense regions is slowed down by the Hubble drag in an
expanding universe \citep{Peebles1993}. Hence, increasing the Hubble
parameter in DE cosmologies causes an increasingly lower growth of
structures. If DE is not the cosmological constant, we can assume
fluctuations in both time and space \citep{Pace2010}, in a similar
fashion to what happens for matter. These fluctuations will
therefore have an impact on the growth of matter perturbation
\citep{Abramo2009b}. This is due to the fact that DE is now not only
important at the background level, but its fluctuations can enhance
or dampen also fluctuations in the dark matter fluid.

 The key parameter to describe perturbations for the DE fluid is the
effective sound speed $c^2_{\rm eff}=\delta p_{\rm
de}/\delta\rho_{\rm de}$. In the context of canonical scalar fields,
$c_{\rm eff}^2=1$ (in units of the light speed, $c$), so that DE
perturbations take place on scales equal to or larger than the
horizon and are totally negligible on smaller scales \citep{Hu1998}.
However, for non-canonical scalar fields, like k-essence models, the
effective sound speed is different from 1, so that we can consider
$c_{\rm eff}^2<1$. In this case, DE perturbations can grow and be
significant also on sub-horizon scales \citep{ArmendarizPicon2001}.
In particular, when $c_s^2\ll 1$, perturbations can grow at the same
rate as matter perturbations \citep{Abramo2009b}. Hence, one can
consider the impact of DE perturbations on the formation of cosmic
structure on sub-horizon scales. In this work we focus on the effect
of DE perturbations on the growth rate of matter in linear regime.

The growth of matter perturbation is described by the growth rate
function $f=d\ln{\delta_{\rm m}}/d\ln{a}$, usually parametrized with
the phenomenological functional form $f=\Omega_{\rm m}^{\gamma}$,
where $\Omega_{\rm m}$ is the fractional energy density of matter in
the universe and $\gamma$ is the so-called growth index
\citep{Peebles1980,Peebles1984,Lahav1991}. The growth index for the
concordance $\Lambda$CDM model is approximately scale independent
and constant with value $\gamma\simeq 0.55$
\citep{Wang1998,Linder2005,Linder2007}. Modified gravity models have
$0.40\leq \gamma\leq 0.43$ at the present time $z=0$
\citep{Tsujikawa2009,Gannouji2009}. In the context of the
Dvali-Gadabadze-Porrati (DGP) braneworld gravity model
\citep{Dvali2000}, the growth index is $\gamma\simeq 11/16 \simeq
0.68$ \citep{Linder2007}. Note, however, that this model has
essentially been ruled out by observations, e.g. by the WiggleZ Dark
Energy survey \citep{Shi2012,Lombriser2013}.

In the framework of General Relativity with dark energy
characterised by a constant equation-of-state parameter, the growth
index $\gamma$ is theoretically approximated by
$\gamma\simeq\frac{3(w_{\rm de}-1)}{6w_{\rm de}-5}$
\citep{Silveira1994,Wang1998,Linder2004,Linder2007,Nesseris2008,Lee2010b}
which explicitly reduces to the well known value
$\gamma_{\Lambda}\simeq 6/11$ in the traditional $\Lambda$CDM
cosmology with $w_{\Lambda}=-1$. Observationally, the growth rate
$f(z)$ has been used to constrain DE models and the growth index
$\gamma$ \citep{Nesseris2008,Cai2012}. On the basis of observations,
a wide range of values for
$\gamma=(0.58-0.67)^{+0.11+0.20}_{-0.11-0.17}$ has been obtained
\citep{Nesseris2008,Guzzo2008,Dossett2010,Samushia2012,Hudson2012}.

The weakness in using the observed values $f_{obs}(z)$ of the growth
rate is due to the fact that it is model dependent, so that they can
only be used to test the consistency of the $\Lambda$CDM model.
Here, we rather prefer to use the quantity $f(z)\sigma_8(z)$, where
$\sigma_8(z)$ is the time-dependant rms amplitude of the overdensity
$\delta_{\rm m}$ at the comoving scale of $8~h^{-1}$~Mpc. The
advantage is that this quantity is almost model independent and
therefore suits better in constraining DE models \citep{Song2009}.
Using the observational growth data $f(z)\sigma_8(z)$, we will
constraint CDE models with equation of state $w_{\rm de}$. As
mentioned before, in CDE models the sound horizon of DE
perturbations is well within the Hubble scale $H^{-1}$ and DE
clusters on scales outside its sound horizon and smaller than the
Hubble length \citep{Appleby2013}. In the same way, cold dark matter
clumps on scales greater than its own Jeans scale. Instead for
canonical DE models, the sound horizon is equal to or larger than
the Hubble length and DE perturbations cannot grow on sub-Horizon
scales. In this work we explore the signature of CDE models on the
growth index of cosmic structures. We put constraints on the growth
index $\gamma$ for a given CDE model and compare the results with
the standard $\Lambda$CDM scenario. We will show that in CDE models,
$\gamma$ may have substantial variations with cosmic redshift, so
that the usual constant growth index parametrization is in general
not suitable in these cases. Due to the inaccuracy of a constant
$\gamma$ parametrization in CDE, we use a general redshift dependent
exponential parametrization and obtain the best fit values of the
parameters.

The plan of the paper is as follows. In Sect.~\ref{sect:gamma} we
present the coupled differential equations governing the
perturbations of matter and DE and argue about the growth index in
such a system. The effect of DE perturbations on the growth index is
discussed. In Sect.~\ref{sect:constraints} we constrain the EoS
parameter of DE for two different models, with constant and
time-varying equation of state, by means of the $\chi^2$ analysis.
We then evaluate $f(z)\sigma_8(z)$ and compare our theoretical
prediction with observational data. Our results show that
perturbations in the DE component give a better fit to observational
data with respect to the $\Lambda$CDM model. We finally conclude in
Sect.~\ref{sect:conclusions}.

\section{Growth index and dark energy perturbation}\label{sect:gamma}
In an Einstein-de Sitter (EdS) universe, in the linear perturbation theory for scales much smaller than the Hubble
radius, the following differential equation describes the evolution of the perturbations of pressureless matter
overdensities $\delta_{\rm m}=\delta\rho_{\rm m}/\rho_{\rm m}$
\citep{Linder2007,Linder2005,Huterer2007,Wang1998,Nesseris2008,Chiba2007}:

\begin{equation}\label{eqn:deltam}
 \ddot{\delta}_{\rm m}+2H\dot{\delta}_{\rm m}-4\pi G\rho_{\rm m}\delta_{\rm m}=0.
\end{equation}

It is well known that Eq.~(\ref{eqn:deltam}) is still valid for DE cosmologies where its effects are only at the
background level. In this case the DE fluid only modifies the Hubble function H. This is not the case any more when DE
perturbations kick in.

In the case of CDE models, we use the following set of linear coupled differential equations obtained in the
framework of the spherical collapse model with the usual top-hat approximation by \cite{Abramo2009b}:

\begin{eqnarray}
&&\ddot{\delta}_{\rm m}+2H\dot{\delta}_{\rm m}=\nonumber\\
&&\quad\frac{3H^2}{2}\left[\Omega_{\rm m}\delta_{\rm m}+\Omega_{\rm de}\delta_{\rm de}(1+3w_{\rm de})\right]\;,
\label{eqn:deltadm1}\\
&&\ddot{\delta}_{\rm de}+\left(2H-\frac{\dot{w}_{\rm de}}{1+w_{\rm de}}\right)\dot{\delta}_{\rm de}=\nonumber \\
&&\quad\frac{3}{2}H^2(1+w_{\rm de})\left[\Omega_{\rm m}\delta_{\rm m}+
\Omega_{\rm de}\delta_{\rm de}(1+3w_{\rm de})\right]\;,\label{eqn:deltadm2}
\end{eqnarray}
where overdot denotes the derivative with respect to cosmic time $t$, $\delta_{\rm de}$ and $w_{\rm de}$ indicate the
perturbations and the EoS parameter of DE, respectively. Under the Top-Hat approximation, $\delta_{\rm m}$ and
$\delta_{\rm de}$ are uniform inside the perturbed region and therefore are a function of cosmic time only.
It should be emphasized that perturbation theory based on the Top-Hat approximation is fully consistent with GR and
Pseudo-Newtonian cosmology in the linear regime \citep{Abramo2009b}. Indeed, \cite{Abramo2009b} showed that the growing
modes of perturbations in GR and in the Pseudo-Newtonian formalism are the same and differences appear only in the
decaying mode. Using the relation $\frac{d}{dt}=aH\frac{d}{da}$, Eqs.~(\ref{eqn:deltadm1} \&
\ref{eqn:deltadm2}) can be written as
\begin{eqnarray}
&& a^2\delta_{\rm m}^{\prime\prime}+\frac{3}{2}a(1-\Omega_{\rm de}
w_{\rm de})\delta_{\rm m}^{\prime}= \nonumber\\
&&\quad
\frac{3}{2}\left[\Omega_{\rm m}\delta_{\rm m}+\Omega_{\rm de}\delta_{\rm de}(1+3w_{\rm de})\right]\;,\label{eqn:deltama}\\
\nonumber && a^2\delta_{\rm
de}^{\prime\prime}+\frac{3}{2}a\left(1-\Omega_{\rm de}w_{\rm de}
-\frac{aw_{\rm de}^{\prime}}{1+w_{\rm de}}\right)\delta_{\rm
de}^{\prime}=
\nonumber\\
&&\quad \frac{3}{2}(1+w_{\rm de})\left[\Omega_{\rm m}\delta_{\rm
m}+\Omega_{\rm de}\delta_{\rm de}(1+3w_{\rm de})\right]\;,
\label{eqn:deltadma}
\end{eqnarray}
where $a$ is the scale factor and primes denote the derivative with
respect to $a$. To derive Eqs.~(\ref{eqn:deltadm1}
and~\ref{eqn:deltadm2}), the EoS parameter of DE, $w_{\rm de}$ is
assumed to be the same inside and outside the perturbed region. We
therefore note that it is possible to derive this set of equations
simply by modifying the source term in Poisson equation (Eq.~4) in
\cite{Pace2010}. To evaluate the impact of DE perturbations as well
as the EoS parameter $w_{\rm de}$ on the growth rate of structures
in CDE models, we solve Eqs.~(\ref{eqn:deltama}
and~\ref{eqn:deltadma}) numerically and obtain $\delta_{\rm m}(z)$
and $\delta_{\rm de}(z)$ where $z$ is the cosmic redshift related to
the scale factor by $z=1/a-1$.

\begin{figure}[ht]
 \centering
 \includegraphics[width=8cm]{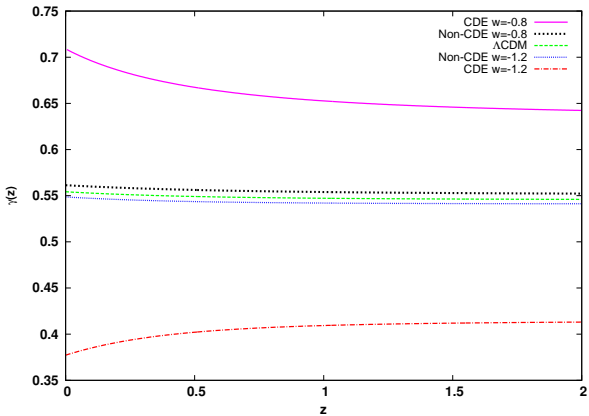}
 \caption{The variation of the growth index $\gamma(z)$ as a function of the cosmic redshift $z$ for different CDE
 and Non-CDE models. The solid violet (black short-dashed) curve represents the clustering (non-clustering)
 quintessence model, the dashed green curve the reference $\Lambda$CDM model, while the blue dotted (red dot-dashed)
 curve stands for the clustering (non-clustering) phantom dark energy model.}
\label{fig:gama}
\end{figure}

For simplicity, in this section we choose a constant EoS parameter
for the dark energy models investigated. We analyse two different
regimes: the phantom one ($w_{\rm de}=-1.2$) and the quintessence
one ($w_{\rm de}=-0.8$). A more general dark energy model with time
varying EoS parameter $w_{\rm de}(z)$ will be discussed in next
section. We chose the following adiabatic initial conditions at
early times ($z_{\rm i}=1000$): $\delta_{\rm m}(z_{\rm i})\simeq
1.4\times 10^{-4}$, $\delta_m^{\prime}(z_{\rm i})\simeq -\delta_{\rm
m}(z_{\rm i})/(1+z_{\rm i})$ and $\delta_{\rm de}(z_{\rm
i})=\frac{(1+w_{\rm de})}{1+3w_{\rm de}}\delta_{\rm m}(z_{\rm i})$,
$\delta_{\rm de}^{\prime}(z_{\rm i})\simeq-\delta_{\rm de}(z_{\rm
i})/(1+z_{\rm i})$ \citep{Spergel2007}. To solve the equations, we
set the present values for the matter and the DE density parameters
as $\Omega_{\rm m,0}=0.3175$ and $\Omega_{\rm de,0}=0.6825$ provided
by the WMAP experiments \citep{Komatsu2011}. By calculating
$\delta_{\rm m}(z)$ and using the definition $f(z)=d\ln{\delta_{\rm
m}(z)}/d\ln{a}$ together with the parametrization $f(z)=\Omega_{\rm
m}(z)^{\gamma(z)}$, we obtain the growth index $\gamma(z)$ from the
coupled Eqs.~(\ref{eqn:deltama} \&~\ref{eqn:deltadma}). In
Fig.~(\ref{fig:gama}) we show the evolution of $\gamma(z)$ as a
function of redshift $z$ for both phantom $w_{\rm de}=-1.2$ and
non-phantom $w_{\rm de}=-0.8$ EoS parameters. The concordance
$\Lambda$CDM cosmology is approximately constant, $\gamma\simeq0.55$
(dashed green line) as expected. Analogously to the $\Lambda$CDM
model, the growth index in non-clustering dark energy (Non-CDE)
models is largely redshift-independent, both for phantom (blue
dotted line) and quintessence (black short-dashed line) regimes. We
also see that differences from the $\Lambda$CDM model are negligible
in Non-CDE models. However, in CDE models, where $\delta_{\rm
de}(z)\neq 0$ and non-negligible, the growth index changes with
redshift and can become as high as $\gamma(z=0)\simeq 0.71$ for the
quintessence case (violet solid curve) and as low as
$\gamma(z=0)\simeq 0.38$ in the phantom case (dot-dashed red curve).
It is also possible to appreciate the large deviation of $\gamma(z)$
in CDE models from the concordance $\Lambda$CDM value $0.55$ as
well. Here we conclude that the impact of DE perturbations on the
growth rate of structures is comparable to the influence of the EoS
parameter. Moreover, $\gamma(z)$ is larger (smaller) than
$\Lambda$CDM value $0.55$ in the case of quintessence (phantom) EoS
parameter.

\section{Observational constraint on the growth index in CDE models}\label{sect:constraints}
In this section we use the current observational growth data
reported in table~(\ref{tab:fsigma8data}) to find the best fit value
for the parameters of the CDE models here studied. We first consider
the CDE model with constant EoS parameter, namely the wCDE model. We
then assume a time varying EoS parameter using the
Chevallier-Polarski-Linder (CPL) parametrization
\citep{Chevallier2001,Linder2003}
\begin{equation}\label{eqn:cpl}
w_{\rm de}(z)=w_0+w_1(\frac{z}{1+z})\;,
\end{equation}
the so-called w(t)CDE model. In the wCDE model, there is one free parameter (the equation of state $w_{\rm de}$) while
for the w(t)CDE model we have two free parameters $(w_0,w_1)$. Considering these two models, we solve the system of
Eqs.~(\ref{eqn:deltama} \&~\ref{eqn:deltadma}) numerically to obtain the theoretical value for growth factor
$f(z)_{\rm the}$, where the subscript \emph{"the"} indicates the theoretical value.\\
We also calculate $\sigma_8(z)_{\rm the}$ as
\begin{eqnarray}
 \sigma_{8}(z=0) & = & \frac{\delta_{\rm m}(z=0)}{\delta_{{\rm m},\Lambda}(z=0)}\sigma_{8,\Lambda}(z=0)\;,\\
 \sigma_8(z)& = & D_{+}(z)\sigma_{8}(z=0)\;,
\end{eqnarray}
where $D_{+}(z)=\delta_{\rm m}(z)/\delta_{\rm m}(z=0)$ is the growth factor. We assume
$\sigma_{8,\Lambda}(z=0)=0.811$, in agreement with the WMAP-7 results \citep{Komatsu2011}.

We also calculate the least square parameter $\chi^2$ according to the definition
\begin{equation}\label{eqn:xi2}
 \chi^2=\sum_{i}\frac{([f(z)\sigma_8(z_i)]_{\rm obs}-[f(z)\sigma_8(z_i)]_{\rm the})^2}{\sigma_{i}^2}\;,
\end{equation}
where the subscript \emph{"obs"} stands for the observational value and $\sigma_i$ is the uncertainty of the
observational data. Currently there are only 10 data points for $f(z)\sigma_8$ in the redshift range $0.067\leq
z\leq0.80$ as shown in table~(\ref{tab:fsigma8data}).

\begin{figure}[!htb]
 \centering
 \includegraphics[width=8cm]{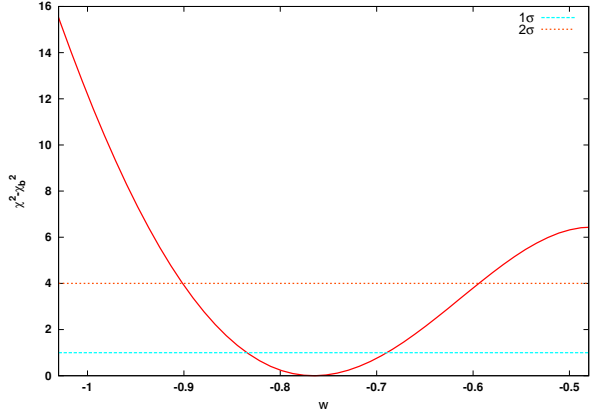}
 \caption{Variation of $\chi^2-\chi^2_{\rm best}$ as a function of $w$ for the wCDE models. The two horizontal lines
 show the $1-\sigma$ and the $2-\sigma$ confidence levels.}
 \label{fig:xi2}
\end{figure}

\begin{figure}[!htb]
 \centering
 \includegraphics[width=8cm]{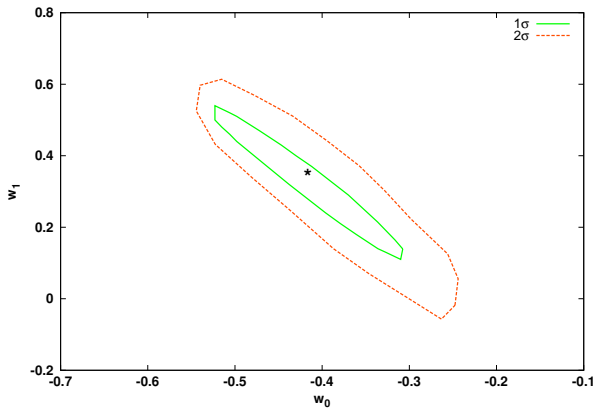}
 \caption{Best parameter values and $1-\sigma$ and $2-\sigma$ confidence levels for the w(t)CDM model. The location of
 the minimum $\chi^2$ is shown a star.}
 \label{fig:xi2-mod2}
\end{figure}

\begin{figure}[!htb]
 \centering
 \includegraphics[width=8cm]{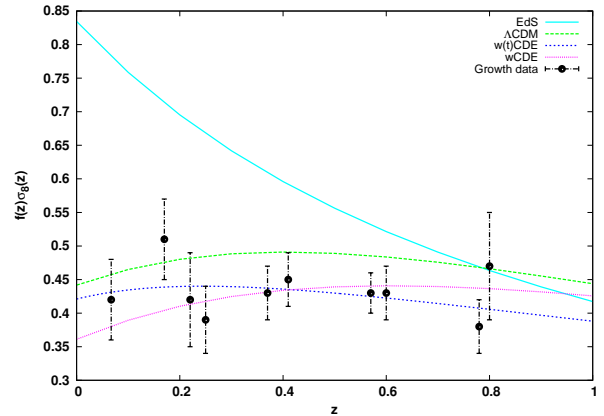}
 \caption{The observed and the theoretical evolution of $f(z)\sigma_8$. The solid cyan curve stands for
the Einstein-de Sitter (EdS) model, the green-dashed one indicates the $\Lambda$CDM model, the pink and blue-dotted
lines represent the wCDE and the w(t)CDE models, respectively. Circles with error-bars are the observed data points
from table~(\ref{tab:fsigma8data}).}
\label{fig:exp-data}
\end{figure}

In Fig.~(\ref{fig:xi2}) we show the evolution of $\chi^2(w)$ as a
function of the dark energy equation of state $w_{\rm de}$ for the
wCDE models. Here we set the best value of $\chi^2$ to zero and show
the $1-\sigma$ and $2-\sigma$ confidence levels by a cyan and an
orange line, respectively. For the wCDE model we obtain the
following best fit value for the EoS parameter: $w_{\rm
de}=-0.764^{+0.073+0.169}_{-0.072-0.136}$ with a $\chi^2_{\rm
best}=6.39$, while for w(t)CDE model we obtain
$w_0=-0.439^{+0.078+0.109}_{-0.132-0.192}$,$w_1=0.375^{+0.175+0.239}_{-0.275-0.136}$
with a $\chi^2_{\rm best}=4.05$. In the case of the concordance
$\Lambda$CDM model we find $\chi^2_{\rm best}=18.58$ and for
comparison we obtain a $\chi^2_{\rm best}=147.09$ for the EdS model.
The results of the $\chi^2$ analysis for the w(t)CDE model is
presented in Fig.~(\ref{fig:xi2-mod2}). As mentioned before, in this
case there are two free parameters $w_0$ and $w_1$ to be
constrained. The star symbol shows the location of the best fit
value and the inner and outer contours represent the $1-\sigma$ and
$2-\sigma$ confidence levels, respectively. In
Fig.~(\ref{fig:exp-data}) we instead present results for the
quantity $f(z)\sigma_8(z)$ for both theoretical models (lines) and
observational data (points with error bars).

\begin{center}
\begin{table}[tbh]
\begin{tabular}{ccp{4.5cm}}
\hline\hline z & $f(z)\sigma_8(z)$ & Survey and Refs \\ \hline
$0.067$& $0.42\pm0.06$ & 6dFGRS~\citep{Beutler2012}\\
$0.17$ & $0.51\pm0.06$ & 2dFGRS~\citep{Percival2004}\\
$0.22$ & $0.42\pm0.07$ & WiggleZ~\citep{Blake2011a}\\
$0.25$ & $0.39\pm0.05$ & SDSS-LRG~\citep{Samushia2012}\\
$0.37$ & $0.43\pm0.04$ & SDSS-LRG~\citep{Samushia2012}\\
$0.41$ & $0.45\pm0.04$ & WiggleZ~\citep{Blake2011a}\\
$0.57$ & $0.43\pm0.03$ & BOSS-CMASS~\citep{Reid2012}\\
$0.60$ & $0.43\pm0.04$ & WiggleZ~\citep{Blake2011a}\\
$0.78$ & $0.38\pm0.04$ & WiggleZ~\citep{Blake2011a}\\
$0.80$ & $0.47\pm0.08$ & VIPERS~\citep{delaTorre2013a}\\
\hline
\end{tabular}
\caption{The $f(z)\sigma_8(z)$ data points including their reference and survey.}
\label{tab:fsigma8data}
\end{table}
\end{center}

In order to select the best model among the ones here studied, we
should compute the reduced chi-squared parameter $\chi^2_{\rm
red}=\chi^2_{\rm best}/\nu$, where $\nu$ represents the number of
degrees of freedom and it is given by $\nu=N-n-1$, where $N$ is the
number of data points and $n$ is the number of fitted parameters.
The deviation from $\chi^2_{\rm red}=1$ measures how good the model
is. For all the models $N=10$. For the wCDE model $n=4$
($\Omega_{\rm m, 0}$, $\Omega_{\rm de, 0}$, $\sigma_8$, $w_{\rm
de}$). In the case of w(t)CDE models, $n=5$ ($\Omega_{\rm m, 0}$,
$\Omega_{\rm de, 0}$, $\sigma_8$, $w_0$, $w_1$). For the concordance
$\Lambda$CDM model $n=3$ ($\Omega_{\rm m, 0}$, $\Omega_{\rm de, 0}$,
$\sigma_8$) and for the EdS model $n=1$ ($\sigma_8$). We find the
following values for the reduced chi-squared
$\chi^2_{\nu}=6.39/5=0.79$ for the wCDE models, $\nu=4.05/4=1.01$
for the w(t)CDE models, $\nu=18.58/6=3.09$ for the concordance
$\Lambda$CDM and $\nu=147.09/8=18.38$ for EdS scenarios. We can
explicitly see that the CDE models fit better the data points than
the standard $\Lambda$CDM and EdS scenarios. Moreover, the w(t)CDE
model is the best model with $0.01$ deviation from $\chi^2_{\rm
red}=1$.

\begin{figure}[!htb]
 \centering
 \includegraphics[width=8cm]{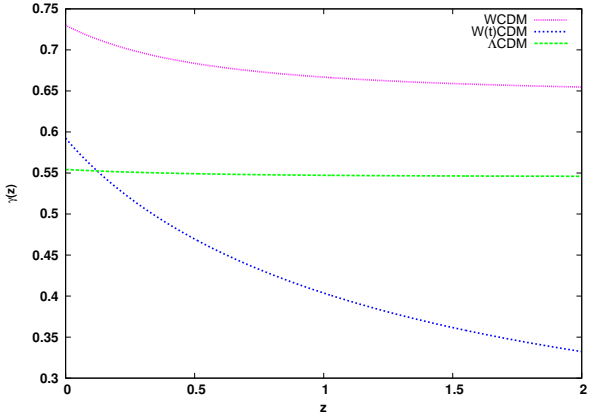}
 \caption{Growth index for different cosmological models on the basis of best fit values of the EoS parameter.}
\label{fig:gama-best}
\end{figure}

We can now evaluate the growth index
$\gamma(z)=\frac{\ln{f(z)}}{\ln{\Omega_{\rm m}}}$ on the basis of
best fit value for $w_{\rm de}$ for both wCDE and w(t)CDE models.
Results are shown in Fig.~(\ref{fig:gama-best}). The $\Lambda$CDM
model is shown by the green dashed line and as expected is
practically a constant. The pink dotted curve represents the wCDE
model ($0.66\leq\gamma(z)\leq 0.73$ in the redshift interval $0\leq
z\leq2$) and is $32\%$ bigger than the present value of the
$\Lambda$CDM model ($\gamma=0.55$). In the case of the w(t)CDE
model, $\gamma(z)$ is $0.33\leq\gamma(z)\leq 0.59$ which is $9\%$
bigger than the present value of the $\Lambda$CDM model (blue
short-dashed curve). In this case $\gamma(z)$ crosses the constant
$\Lambda$CDM line at $z\simeq0.15$ representing a smaller growth
index for $z>0.15$ and a bigger one for $z<0.15$ compared to the
$\Lambda$CDM model. It is important to note that, unlike
$\Lambda$CDM cosmologies, in CDE models the growth index $\gamma$
has a wide range of values, in particular in the case of w(t)CDE
models. It should also be noted that for CDE models, due to the
large variation of $\gamma(z)$ with respect to the redshift $z$, the
constant $\gamma_0$ parametrization is not accurate enough. Hence,
it is worth using a parametrized function of $\gamma(z)$ in order to
discuss the best fit value of the growth index in CDE models. Using
the fact that the growth index should be nearly constant at early
times and change at late times, we apply the following general
exponential parametrization \citep{Dossett2010}:

\begin{equation}\label{eqn:param}
 \gamma(z)=\gamma_{\infty}+\gamma_{\rm b} e^{-z/z_{\rm t}}
\end{equation}
where $z_{\rm t}$ represents the transition redshift from an almost constant $\gamma_{\infty}$ at early times to a
varying growth index $\gamma(z)$ at late times. In the case of the concordance $\Lambda$CDM cosmology, the exponential
parametrization yields the best fit parameters $\gamma_{\infty}=0.5457$, $\gamma_{\rm b}=0.0103$ and $z_{\rm t}=0.61$
\citep{Dossett2010}. In table~\ref{tab:param}, we show the best fit parameters of $\gamma_{\infty}$, $\gamma_{\rm b}$
and $z_{\rm t}$ with the corresponding error-bars for the wCDE and w(t)CDE models. A large value of $\gamma_{\rm b}$
represents a strong variation of $\gamma(z)$ with redshift for w(t)CDE model.

\begin{center}
\begin{table}[tbh]
\begin{tabular}{ccl}
\hline         Model &           wCDM       &      w(t)CDM  \\\hline
$\gamma_{\infty}$    &  $0.648984^{+0.00006}_{-0.00006}$       & $0.253179^{+0.0004126}_{-0.0004126}$ \\
$\gamma_{\rm b}$     & $ 0.0750995^{+0.00021}_{-0.00021}$      & $0.318083^{+ 0.00055}_{- 0.00055}$ \\
$z_{\rm t}$          &  $0.698723^{+ 0.003559 }_{- 0.003559 }$ & $1.39674^{+ 0.006793 }_{- 0.006793 }$ \\
\hline
\end{tabular}
\caption{The best fit parameters for $\gamma(z)$ on the basis of the parametrization of Eq.~(\ref{eqn:param}).}
\label{tab:param}
\end{table}
\end{center}

\section{Conclusions}\label{sect:conclusions}
In this work we studied the evolution of density perturbations of pressureless dark matter and dark energy on
sub-horizon scales by solving the linear equations of structure formation with a density profile assumed to be
described by a Top-Hat profile. We showed that the matter growth index $\gamma$ is strongly affected by perturbations
of the DE component, while for smooth DE models it is almost constant with cosmic time. It has been shown that for
CDE models, the growth index $\gamma$ can be larger (for quintessence EoS $w_{\rm de}>-1$) or smaller (for the phantom
EoS $w_{\rm de}<-1$) than the concordance $\Lambda$CDM predicted value $\gamma_{\Lambda}\simeq0.55$. We performed a
reduced $\chi^2$ minimization analysis between the observations and the theoretical expectations of $f(z)\sigma_8(z)$
and found that for CDE with constant EoS parameter, i.e., wCDE model:
$w_{\rm de}=-0.764^{+0.073+0.169}_{-0.072-0.136}$ with a reduced $\chi^2_{\rm red}=0.79$, while for w(t)CDE model
with a CPL parametrization for $w_{\rm de}(z)$ \citep{Chevallier2001,Linder2003}, we
found:($w_0=-0.439^{+0.078+0.109}_{-0.132-0.192}$,$w_1=0.375^{+0.175+0.239}_{-0.275-0.136}$)
with a $\chi^2_{\rm red}=1.01$. For comparison, the concordance $\Lambda$CDM model has $\chi^2_{red}=3.09$.
Hence the w(t)CDE model fits better to the observational growth data with respect to other models.

We also discussed the growth index $\gamma$ in the framework of
clustering dark energy models and we found a strong variation with
respect to redshift, particularly in the case of the w(t)CDE model.
We showed that for the best fit value of $w_{\rm de}$, the w(t)CDE
model crosses the $\Lambda$CDM solution at $z\simeq0.15$, with lower
values for $z>0.15$ and higher values for $z<0.15$ when compared to
the $\Lambda$CDM result. Finally, we applied the general exponential
parametrization for $\gamma(z)$ in Eq.~(\ref{eqn:param}) and
predicted that for the wCDE model analysed
$\gamma_{\infty}>\gamma_{\infty}^{\Lambda CDM}$, while for the
w(t)CDE model $\gamma_{\infty}<\gamma_{\infty}^{\Lambda CDM}$ (see
table~\ref{tab:param}). Hence the growth index at early times is
larger for wCDE and smaller for w(t)CDE models compare with the
$\Lambda$CDM model. The large value of $\gamma_{\rm b}$ in
table~\ref{tab:param} indicates a strong variation of the growth
index in w(t)CDE cosmologies.

\section*{Acknowledgements}
F. Pace is supported by STFC grant ST/H002774/1.\\
We would like to thank anonymous referee for giving constructive
comments and reading the manuscript.

%\bibliographystyle{spr-mp-nameyear-cnd}
%\bibliography{gamma.bbl}

\begin{thebibliography}{99}

\bibitem [Abramo et al.(2009)]{Abramo2009b}
Abramo, L.R., Batista, R.C., Liberato, L., Rosenfeld, R.: Phys. Rev.
D 79(2), 023516 (2009)


\bibitem [Alam et~al.(2003)]{Alam2003}
Alam, U., Sahni, V., Deep Saini, T., Starobinsky, A.A.: Mon. Not. R.
Astron. Soc. 344, 1057 (2003)


\bibitem [Amendola et~al.( 2007)]{Amendola2007}
Amendola, L., Gannouji, R., Polarski, D., Tsujikawa, S.: Phys. Rev.
D 75(8), 083504 (2007)


\bibitem[Appleby \& Battye(2007)]{Appleby2007}
Appleby, S., Battye, R.: Physics Letters B 654, 7 (2007)


\bibitem[Appleby et~al.(2013)]{Appleby2013}
Appleby, S.A., Linder, E.V., Weller, J.: Phys. Rev. D 88(4), 043526
(2013)


\bibitem[Armendariz-Picon et~al.(2001)]{ArmendarizPicon2001}
Armendariz-Picon, C., Mukhanov, V., Steinhardt, P.J.: Phys. Rev. D
63(10), 103510 (2001)


\bibitem[Bennett et~al.(2003)]{Bennett2003}
Bennett, C.L., Halpern, M., Hinshaw, G., Jarosik, N., Kogut, A.,
Limon, M., Meyer, S.S., et al.: Astrophys. J. Suppl. Ser. 148, 1
(2003)


\bibitem[Beutler et~al.(2012)]{Beutler2012}
Beutler, F., Blake, C., Colless, M., Jones, D.H., Staveley-Smith,
L., Poole, G.B., Campbell, L., et al.: Mon. Not. R. Astron. Soc.
423, 3430 (2012)


\bibitem[Blake et~al.(2011)]{Blake2011a}
Blake, C., Brough, S., Colless, M., Contreras, C., Couch, W., Croom,
S., Davis, T., Drinkwater, M.J., Forster, K., et al.: Mon. Not. R.
Astron. Soc. 415, 2876 (2011)



\bibitem[Cai et~al.(2012)]{Cai2012}
Cai, R.-G., Tuo, Z.-L., Wu, Y.-B., Zhao, Y.-Y.: Phys. Rev. D 86(2),
023511 (2012)


\bibitem[Chevallier \& Polarski(2001)]{Chevallier2001}
Chevallier, M., Polarski, D.: International Journal of Modern
Physics D 10, 213 (2001)


\bibitem[Chiba \& Takahashi(2007)]{Chiba2007}
Chiba, T., Takahashi, R.: Phys. Rev. D 75(10), 101301 (2007)


\bibitem[Cognola et~al.(2008)]{Cognola2008}
Cognola, G., Elizalde, E., Nojiri, S., Odintsov, S.D., Sebastiani,
L., Zerbini, S.: Phys. Rev. D 77(4), 046009 (2008)


\bibitem[Cole et~al.(2005)]{Cole2005}
Cole, S., Percival, W.J., Peacock, J.A., Norberg, P., Baugh, C.M.,
Frenk, C.S., Baldry, I., Bland-Hawthorn, J., Bridges, T., et al.:
Mon. Not. R. Astron. Soc. 362, 505 (2005)


\bibitem[de la Torre et~al.(2013)]{delaTorre2013a}
de la Torre, S., Guzzo, L., Peacock, J.A., Branchini, E., Iovino,
A., Granett, B.R., Abbas, U., Adami, C., Arnouts, S., et al.:
Astron. Astrophys. 557, 54 (2013)


\bibitem[Dossett et~al.(2010)]{Dossett2010}
Dossett, J., Ishak, M., Moldenhauer, J., Gong, Y., Wang, A.: J.
Cosmol. Astropart. Phys. 4, 22 (2010)


\bibitem[Dvali et~al.(2000)]{Dvali2000}
Dvali, G., Gabadadze, G., Porrati, M.: Physics Letters B 485, 208
(2000)


\bibitem[Eisenstein et~al.(2005)]{Eisenstein2005}
Eisenstein, D.J., Zehavi, I., Hogg, D.W., Scoccimarro, R., Blanton,
M.R., Nichol, R.C., Scranton, R., Seo, H.-J., Tegmark, M., Zheng,
Z., et al.: Astrophys. J. 633, 560 (2005)


\bibitem[Gannouji et~al.(2009)]{Gannouji2009}
Gannouji, R., Moraes, B., Polarski, D.: J. Cosmol. Astropart. Phys.
2, 34 (2009)


\bibitem[Guzzo et~al.(2008)]{Guzzo2008}
Guzzo, L., Pierleoni, M., Meneux, B., Branchini, E., Le F\`evre, O.,
Marinoni, C., Garilli, B., Blaizot, J., De Lucia, G., et al.: Nature
451, 541 (2008)


\bibitem[Hawkins et~al.(2003)]{Hawkins2003}
Hawkins, E., Maddox, S., Cole, S., Lahav, O., Madgwick, D.S.,
Norberg, P., Peacock, J.A., Baldry, I.K., Baugh, C.M.,
Bland-Hawthorn, J., et al.: Mon. Not. R. Astron.Soc. 346, 78 (2003)


\bibitem[Hu(1998)]{Hu1998}
Hu, W.: Astrophys. J. 506, 485 (1998)


\bibitem[Hu \& Sawicki(2007)]{Hu2007}
Hu, W., Sawicki, I. : Phys. Rev. D 76(6), 064004 (2007)


\bibitem[Hudson \& Turnbull(2012)]{Hudson2012}
Hudson, M.J., Turnbull, S.J.: Astrophys. J. Lett. 751, 30 (2012)

\bibitem[Huterer \& Linder(2007)]{Huterer2007}
Huterer, D., Linder, E.V.: Phys. Rev. D 75(2), 023519 (2007)


\bibitem[Komatsu et~al.(2011)]{Komatsu2011}
Komatsu, E., Smith, K.M., Dunkley, J., et al.: Astrophys. J. Suppl.
Ser. 192, 18 (2011)


\bibitem[Lahav et~al.(1991)]{Lahav1991}
Lahav, O., Lilje, P.B., Primack, J.R., Rees, M.J.: Mon. Not. R.
Astron. Soc. 251, 128 (1991)


\bibitem[Lazkoz et~al.(2005)]{Lazkoz2005}
Lazkoz, R., Nesseris, S., Perivolaropoulos, L.: J. Cosmol.
Astropart. Phys. 11, 10 (2005)


\bibitem[Lee \& Ng(2010)]{Lee2010b}
Lee, S., Ng, K.-W. : Physics Letters B 688, 1 (2010)


\bibitem[Linder(2003)]{Linder2003}
Linder, E.V.: Physical Review Letters 90(9), 091301 (2003)


\bibitem[Linder(2004)]{Linder2004}
Linder, E.V.: Phys. Rev. D 70(2), 023511 (2004)


\bibitem[Linder(2005)]{Linder2005}
Linder, E.V.: Phys. Rev. D 72(4), 043529 (2005)


\bibitem[Linder(2009)]{Linder2009}
Linder, E.V.: Phys. Rev. D 80(12), 123528 (2009)


\bibitem[Linder \& Cahn(2007)]{Linder2007}
Linder, E.V., Cahn, R.N.: Astroparticle Physics 28, 481 (2007)


\bibitem[Lombriser et~al.(2013)]{Lombriser2013}
Lombriser, L., Yoo, J., Koyama, K.: Phys. Rev. D 87(10), 104019
(2013)


\bibitem[Nesseris \& Perivolaropoulos(2008)]{Nesseris2008}
Nesseris, S., Perivolaropoulos, L.: Phys. Rev. D 77(2), 023504
(2008)


\bibitem[Pace et~al.(2010)]{Pace2010}
Pace, F., Waizmann, J.-C., Bartelmann, M.: Mon. Not. R. Astron. Soc.
406, 1865 (2010)


\bibitem[Peebles \& Ratra(2003)]{Peebles2003}
Peebles, P.J., Ratra, B.: Reviews of Modern Physics 75, 559 (2003)


\bibitem[Peebles(1980)]{Peebles1980}
Peebles, P.J.E.: The Large-scale Structure of the Universe, (1980)



\bibitem[Peebles(1984)]{Peebles1984}
Peebles, P.J.E.: Astrophys. J. 284, 439 (1984)


\bibitem[Peebles(1993)]{Peebles1993}
Peebles, P.J.E.: Principles of Physical Cosmology, (1993)


\bibitem[Percival et~al.(2004)]{Percival2004}
Percival, W.J., Burkey, D., Heavens, A., Taylor, A., Cole, S.,
Peacock, J.A., Baugh, C.M., et al.: Mon. Not. R. Astron. Soc. 353,
1201 (2004)


\bibitem[Perlmutter et~al.(1997)]{Perlmutter1997}
Perlmutter, S., Gabi, S., Goldhaber, G., Goobar, A., Groom, D.E.,
Hook, I.M., Kim, A.G., et al.: Astrophys. J. 483, 565 (1997)
astro-ph/9608192


\bibitem[Perlmutter et~al.(1998)]{Perlmutter1998}
Perlmutter, S., Aldering, G., della Valle, M., Deustua, S., Ellis,
R.S., Fabbro, S., Fruchter, A., Goldhaber, G., et al.P: Nature 391,
51 (1998)



\bibitem[Perlmutter et~al.(1999)]{Perlmutter1999}
Perlmutter, S., Aldering, G., Goldhaber, G., et al.: Astro-phys. J.
517, 565 (1999)



\bibitem[Planck Collaboration et~al.(2013)]{Planck2013_XVI}
Planck Collaboration, Ade, P.A.R., Aghanim, N., Armitage-Caplan, C.,
Arnaud, M., Ashdown, M., Atrio-Barandela, F., Aumont, J.,
Baccigalupi, C., Banday, A.J., et al.: ArXiv e-prints, 1303.5076
(2013) 1303.5076



\bibitem[Reid et~al.(2012)]{Reid2012}
Reid, B.A., Samushia, L., White, M., Percival, W.J., Man-era, M.,
Padmanabhan, N., Ross, A.J., et al.: Mon. Not. R. Astron. Soc. 426,
2719 (2012)


\bibitem[Riess et~al.(2004)]{Riess2004}
Riess, A.G., Strolger, L.-G., Tonry, J., Casertano, S., Fer-guson,
H.C., et al.: Astrophys. J. 607, 665 (2004)



\bibitem[Riess et~al.(2007)]{Riess2007}
Riess, A.G., Strolger, L.-G., Casertano, S., Ferguson, H.C.,
Mobasher, B., Gold, B., Challis, P.J., et al.: Astrophys. J. 659, 98
(2007)


\bibitem[Sahni \& Starobinsky(2000)]{Sahni2000}
Sahni, V., Starobinsky, A.: International Journal of Modern Physics
D 9, 373 (2000)


\bibitem[Sahni et~al.(2003)]{Sahni2003}
Sahni, V., Saini, T.D., Starobinsky, A.A., Alam, U.: Soviet Journal
of Experimental and Theoretical Physics Letters 77, 201 (2003)


\bibitem[Samushia et~al.(2012)]{Samushia2012}
Samushia, L., Percival, W.J., Raccanelli, A.: Mon. Not. R. Astron.
Soc. 420, 2102 (2012)


\bibitem[Shi et~al.(2012)]{Shi2012}
Shi, K., Huang, Y.F., Lu, T.: Mon. Not. R. Astron. Soc. 426, 2452
(2012)


\bibitem[Silveira \& Waga(1994)]{Silveira1994}
Silveira, V., Waga, I.: Phys. Rev. D 50, 4890 (1994)


\bibitem[Song \& Percival(2009)]{Song2009}
Song, Y.-S., Percival, W.J.: J. Cosmol. Astropart. Phys. 10, 4
(2009)


\bibitem[Spergel et~al.(2003)]{Spergel2003}
Spergel, D.N., Verde, L., Peiris, H.V., Komatsu, E., Nolta, M.R.,
Bennett, C.L., Halpern, M., Hinshaw, G., Jarosik, N., Kogut, A.,
Limon, M., Meyer, S.S., Page, L., Tucker, G.S., Weiland, J.L.,
Wollack, E., Wright, E.L.: Astro-phys. J. Suppl. Ser. 148, 175
(2003)


\bibitem[Spergel et~al.(2007)]{Spergel2007}
Spergel, D.N., Bean, R., Dor\ ́e, O., Nolta, M.R., Bennett, C.L.,
Dunkley, J., Hinshaw, G., Jarosik, N., Komatsu, E., Page, L., et
al.: Astrophys. J. Suppl. Ser. 170, 377 (2007)


\bibitem[Starobinsky(2007)]{Starobinsky2007}
Starobinsky, A.A.: Soviet Journal of Experimental and Theoretical
Physics Letters 86, 157 (2007)


\bibitem[Tegmark et~al.(2004)]{Tegmark2004a}
Tegmark, M., Strauss, M.A., Blanton, M.R., Abazajian, K., Dodelson,
S., Sandvik, H., Wang, X., Weinberg, D.H., Zehavi, I., Bahcall,
N.A., et al.: Phys. Rev. D 69(10), 103501 (2004)



\bibitem[Tsujikawa et~al.(2008)]{Tsujikawa2008}
Tsujikawa, S., Uddin, K., Mizuno, S., Tavakol, R., Yokoyama, J.:
Phys. Rev. D 77(10), 103009 (2008)


\bibitem[Tsujikawa et~al.(2009)]{Tsujikawa2009}
Tsujikawa, S., Gannouji, R., Moraes, B., Polarski, D.: Phys. Rev. D
80(8), 084044 (2009)


\bibitem[Wang \& Steinhardt(1998)]{Wang1998}
Wang, L., Steinhardt, P.J.: Astrophys. J. 508, 483 (1998)


\bibitem[Weinberg(1989)]{Weinberg1989}
Weinberg, S.: Reviews of Modern Physics 61, 1 (1989)


\end{thebibliography}

\end{document}